\documentclass{PoS}
\usepackage{subcaption}
\title{Reweighting Lefschetz Thimbles}

\ShortTitle{Reweighting Lefschetz Thimbles}

\author{Stefan Bl\"ucher\\
        Universit\"at Heidelberg, Heidelberg, Germany\\
        E-mail: \email{bluecher@stud.uni-heidelberg.de}}

\author{Jan M.~Pawlowski\\
        Institut f\"ur Theoretische Physik, Universit\"at Heidelberg, Heidelberg, Germany\\
        E-mail: \email{j.pawlowski@thphys.uni-heidelberg.de}}

\author{Manuel Scherzer\\
        Institut f\"ur Theoretische Physik, Universit\"at Heidelberg, Heidelberg, Germany\\
        E-mail: \email{scherzer@thphys.uni-heidelberg.de}}

\author{Mike Schlosser\\
        Universit\"at Heidelberg, Heidelberg, Germany\\
        E-mail: \email{m.schlosser@stud.uni-heidelberg.de}}

\author{Ion-Olimpiu Stamatescu\\
        Institut f\"ur Theoretische Physik, Universit\"at Heidelberg, Heidelberg, Germany\\
        E-mail: \email{i.o.stamatescu@thphys.uni-heidelberg.de}}

\author{Sebastian Syrkowski\\
        Universit\"at Heidelberg, Heidelberg, Germany\\
        E-mail: \email{syrkowski@stud.uni-heidelberg.de}}

\author{\underline{\speaker{Felix P.~G.~Ziegler}}\\
        Institut f\"ur Theoretische Physik, Universit\"at Heidelberg, Heidelberg, Germany\\
        E-mail: \email{f.ziegler@thphys.uni-heidelberg.de}}

\abstract{We present a novel reweighting technique to calculate the relative weights 
          in the Lefschetz thimble decomposition of a path integral. 
          Our method is put to work using a $U(1)$ one-link model providing for a suitable testing 
          ground and sharing many features with realistic gauge theories with fermions at 
          finite density. We discuss prospects and future challenges to our method.}

\FullConference{The 36th Annual International Symposium on Lattice Field Theory - LATTICE2018\\
		22-28 July, 2018\\
		Michigan State University, East Lansing, Michigan, USA.}

\begin{document}

\section{Introduction}

\noindent The exploration of the phase diagram of Quantum Chromodynamics (QCD) 
provides a formidable challenge to lattice simulations due to the sign problem. 
The latter arises for finite chemical potential $\mu$ rendering the Euclidean 
action complex. Consequently, 
importance sampling based methods relying on a positive definite Boltzmann 
measure become inapplicable. In recent years, a plethora of contenders have been 
proposed towards mitigating the sign problem, see e.g.~\cite{deForcrand:2010ys}.
So far none of the method 
has been able to produce reliable results for $\mu / T \geq 1$.
Amongst those, promising candidates operating on complex manifolds are the 
Complex Langevin evolution \cite{Aarts:2008rr, Aarts:2017hqp} the Lefschetz thimble method 
\cite{Cristoforetti:2012su} and flowed manifolds \cite{Alexandru:2015sua}. 
Recently, methods to find an optimal complex integration manifold where the sign problem is mild 
have been proposed in \cite{Mori:2017pne, Alexandru:2018fqp, Bursa:2018ykf}.
In this work we consider the Lefschetz thimbles which have been applied recently to 
one-dimensional QCD in \cite{Schmidt:2017gvu, DiRenzo:2017igr}.
We propose algorithmic improvements %for Monte Carlo simulations on Lefschetz thimbles
addressing the determination of the relative weights of the contributing thimbles
to the partition function. 
%This problem arises from the decomposition of the original 
%path integral into a sum of integrals over multiple thimbles. 
%We show how this can be 
%dealt with via 
To this end we present a novel reweighting 
procedure which can be generalized to higher dimensional theories.
Our method is demonstrated by using a $U(1)$ one-link model representing a 
suitable testing ground towards gauge theories with fermions at finite density.

%The advantage is that
%the thimble structure is straightforward and thus provides a useful testbed to probe novel
%algorithms.

\section{Lefschetz thimbles}
\label{sec:LT-intro}
\noindent For simplicity we consider a complex action $S(x)$ of a real variable 
that we extend to the complex plane $x \to z = x + i y$.
The Lefschetz thimbles \cite{Cristoforetti:2012su, Witten:2010cx} 
are the paths of steepest descent $D_{\sigma} \in \mathbb{C}$
of the real part of the action and are obtained from the flow equation
\begin{equation}
\frac{\partial z}{\partial \tau}=-\overline{\frac{\partial S}{\partial z}}\,.
\label{eq:steep-desc}
\end{equation}
The thimbles end in the critical points $z_{\sigma} \in \mathbb{C}$ of the action
which are the stationary solutions to the right-hand side of (\ref{eq:steep-desc}).
%The critical points $z_{\sigma}$ of $S(z)$ are determined from
%\begin{equation}
%\left. \frac{\partial S}{\partial z}\right|_{z=z_{\sigma}}=0\,.
%\end{equation}
%For each critical point of $z_\sigma$ the path $D_{\sigma}\subset \mathbb{C}$ 
%solving the steepest descent equation  
%and ending in $z_\sigma$ is called a Lefschetz thimble. 
Along the thimble the real part of the action decreases and has its
minimum at the critical point $z_{\sigma}$. Importantly, the imaginary
part of the action is constant along each thimble which can ameliorate
the sign problem.  The partition function can be rewritten as an
integral over the union of the thimbles whose anti-thimbles 
(path of steepest ascent of $\mathrm{Re}[S]$) cross the
real axis.  If the critical points are non-degenerate this holds by
continuously deforming the original real integration contour into the
thimbles.  Hence, for the partition function we have
\begin{equation}
  Z=\int_{I} dx\,e^{-S(x)}=\sum_{\sigma}n_{\sigma}
                e^{-i\mathrm{Im}\left[S\left(z_{\sigma}\right)\right]}
                \underbrace{\int_{D_{\sigma}}dz\, 
                e^{-\mathrm{Re}\left[S\left(z\right)\right]}}_{=: Z_{\sigma}}\,.
\label{eq:Thimble_basics}
\end{equation}
Here, $I \subset \mathbb{R}$ denotes the original real integration domain. 
$n_{\sigma}$ counts the number of intersections 
of a given anti-thimble with $I$.  
The latter is also called the unstable thimble since the action is not bounded
from below.
Observables are obtained from the expression
\begin{equation}
	\langle \mathcal{O} \rangle = \frac{1}{Z}
	\sum_{\sigma}n_{\sigma} e^{-i\mathrm{Im}\left[S\left(z_{\sigma}\right)\right]}
	\int_{D_{\sigma}}dz\,\mathcal{O}e^{-\mathrm{Re}\left[S\left(z\right)\right]}
	= \frac{\sum_{\sigma} n_{\sigma} e^{-i\mathrm{Im}S\left(z_{\sigma}\right)}Z_{\sigma} 
    \langle\mathcal{O}\rangle_{\sigma}}{\sum_{\sigma} n_{\sigma} e^{-i\mathrm{Im}S\left(z_{\sigma}\right)}Z_{\sigma}}\,,
\label{eq:th-decomp-obs}
\end{equation}
where we have defined 
\begin{equation}
\langle \mathcal{O} \rangle_{\sigma} := \frac{1}{Z_\sigma}
\int_{D_\sigma} dz e^{-\mathrm{Re}[S(z)]} \mathcal{O}\,.
\label{eq:obs-single-thimble}
\end{equation}
This approach has two practical challenges:
\begin{enumerate}
\item Monte Carlo sampling on the (main) contributing thimbles based on finding a numerical 
      parametrization implicitly or explicitly.
\item In most realistic theories multiple thimbles contribute to the partition function. Hence, 
      we need to determine the relative weights
      weights $Z_{\rho}/Z_{\sigma}$ where $\rho \neq \sigma$ in (\ref{eq:th-decomp-obs}). 
\end{enumerate}
Numerical cost and complexity of the above tasks increase 
with the dimensionality of the considered theory.
So far there has not been a general solution capturing both problems.
However, the first difficulty can be tackled by the holomorphic flow equations
continuously deforming the original integration contour towards 
the thimbles \cite{Alexandru:2015xva}. Another approach addresses
the second problem by means of a semi-classical approximation \cite{DiRenzo:2017omx}.
Methods tackling both problems 
have been proposed in \cite{Bluecher:2018sgj}.
In this presentation, we focus only on the second 
problem and present a technique to compute the relative weights. 
%demonstrated 
%to  where we discuss some ideas and their feasibility 
%regarding the first problem. 
%We demonstrate our technique
%by means of simple one-dimensional integrals.

\section{Simulation method on Lefschetz thimbles}
\label{sec:monte_carlo}
\noindent In this section we present an algorithm for a Monte Carlo simulation on
Lefschetz thimbles, assumed that we know a parametrization of all contributing thimbles. 
For the discussion of finding a numerical 
parametrization in simple models we refer the reader to \cite{Bluecher:2018sgj}. 
The method is divided up into two steps. First we discuss how to compute 
the expectation value of a given observable on a single thimble
$D_{\sigma}$ according to (\ref{eq:obs-single-thimble}). In the second step
we demonstrate how the ratios of partition functions determining the 
relative weights of the thimbles can be computed within the Monte Carlo simulation. 
%With both steps completed (\ref{eq:th-decomp-obs}) can be calculated.
For the first step let 
$[a, b] \subset \mathbb{R} \to D_{\sigma}: \tau \mapsto z(\tau)$
be a (numerical) parametrization 
of the thimble $D_{\sigma}$. Its associated 
partition function reads
\begin{equation}
  Z_{\sigma} = \int_{D_{\sigma}} dz\,e^{-\mathrm{Re}\left[S(z)\right]} = \int_{a}^{b}  d\tau\,
  e^{-\mathrm{Re}\left[S_\sigma(\tau)\right]} J_\sigma(z(\tau))\,.
\label{eq:jacobian}
\end{equation}
Here $S_{\sigma}(\tau)=S(z(\tau))$ denotes the action evaluated on 
$D_{\sigma}$. Moreover, the complex Jacobian 
$J_{\sigma}(\tau) := \partial z(\tau) / \partial \tau$ on $D_\sigma$ has been introduced.
We define   
\begin{equation}
\langle \mathcal{O} \rangle_{\sigma}^{r} = 
\frac{1}{Z_{\sigma}^r}\int_{a}^{b} d\tau\,e^{-\mathrm{Re}\left[S_\sigma\left(\tau\right) \right]} \mathcal{O}\, ,
\label{eq:real-sub-ev}
\end{equation}
as well as
\begin{equation}
Z_{\sigma}^{r}= \int_{a}^{b} d\tau\,e^{-\mathrm{Re}\left[S_\sigma\left(\tau\right)\right]}\,.
\label{eq:single-th-rpf}
\end{equation}
%The right-hand side of (\ref{eq:jacobian}) can be computed by sampling $\tau$ distributed according 
%to $e^{-\mathrm{Re}[S_{\sigma}(\tau)]} / Z_{\sigma}^r$. 
%\begin{equation}
%p_\sigma(\tau)= \frac{e^{-\mathrm{Re}\left[S_{\sigma}(\tau)\right]}}{Z^r_\sigma}\,.
%\label{eq:distribution}
%\end{equation}
Thus, to calculate an observable on a single thimble we sample $\tau$
distributed according 
to $e^{-\mathrm{Re}[S_{\sigma}(\tau)]} / Z_{\sigma}^r$ and take into  
account the Jacobian via conventional reweighting 
\begin{equation}
\langle \mathcal{O} \rangle_{\sigma} = \frac{\langle \mathcal{O} J_{\sigma}\rangle_{\sigma}^r}{\langle J_{\sigma}\rangle_{\sigma}^r}\,.
\label{eq:single-thimble-obs-rew}
\end{equation}
The phase of the Jacobian gives rise to the so-called residual sign problem.
The expectation value of an observable reads
% in the thimble decomposition (\ref{eq:th-decomp-obs}) 
\begin{equation}
\left<\mathcal{O}\right>=\frac{\sum_{\sigma} n_{\sigma} 
e^{-i\mathrm{Im}\left[S\left(z_\sigma\right)\right]}Z^{r}_{\sigma} \left<
\mathcal{O}\,J_\sigma \right>_{\sigma}^{r}}{\sum_{\sigma} n_{\sigma} 
e^{-i\mathrm{Im}\left[S\left(z_\sigma\right)\right]}Z^{r}_{\sigma} \left<J_\sigma\right>_{\sigma}^{r}}\,.
\label{eq:obs_full}
\end{equation}
It is obvious that in the case 
of a single contributing thimble (\ref{eq:obs_full})  
reduces to (\ref{eq:single-thimble-obs-rew}).
In the second step we capture 
the weights from the partition functions $Z_\sigma^r$
from within the Monte Carlo simulation.
For simplicity we consider (\ref{eq:obs_full}) with only two contributing thimbles, 
i.e.~$\sigma = 1,2$.
Without loss of generality we choose the second thimble and factor out $Z_2^r$
in (\ref{eq:obs_full}). This thimble is 
refered to as the "master thimble".
Hence, we are left with computing the ratio of the 
partition functions $Z_1^r$ and $Z_2^r$ which can be straightforwardly written in terms
of the expectation value (\ref{eq:real-sub-ev}) of the ratio of the associated Boltzmann factors
\begin{equation}
\frac{Z_{1}^r}{Z_{2}^r}= 
\left<e^{\mathrm{Re}\left[S_2-S_1\right]}\right>_{2}^{r}\,.
\label{eq:ratio}
\end{equation}
This means that we determine the relative weights in (\ref{eq:obs_full})
by reweighting with respect to the master thimble.
We remark that for (\ref{eq:ratio}) to hold (a) the integrals over the thimbles 
have the same parameter interval $[a, b]$
and (b) the parameters $\tau$
on both thimbles must be identified. In case the latter does not hold 
we need to take into account an additional Jacobian. For the example integrals 
considered in this work (a) can be realized 
by suitable variable transformations and (b) can be achieved by normalizing 
the steepest descent equation when determining the numerical parametrizations.
For further details, see \cite{Bluecher:2018sgj}
and appendices A and B therein. 
Thus, we can compute the relative weights of the contributing thimbles (\ref{eq:ratio})
within the Monte Carlo simulation by reweighting with respect to the master thimble which 
completes the list of ingredients for (\ref{eq:obs_full}).
The reweighting method might also be useful in addressing higher dimensional integrals for field theories
since it does not rely on an explicit (numerical) parametrization of the thimble.

\section{Numerical results}
\begin{figure}[t]
	\begin{subfigure}[c]{.48\textwidth}
		\includegraphics[width=.9\textwidth]{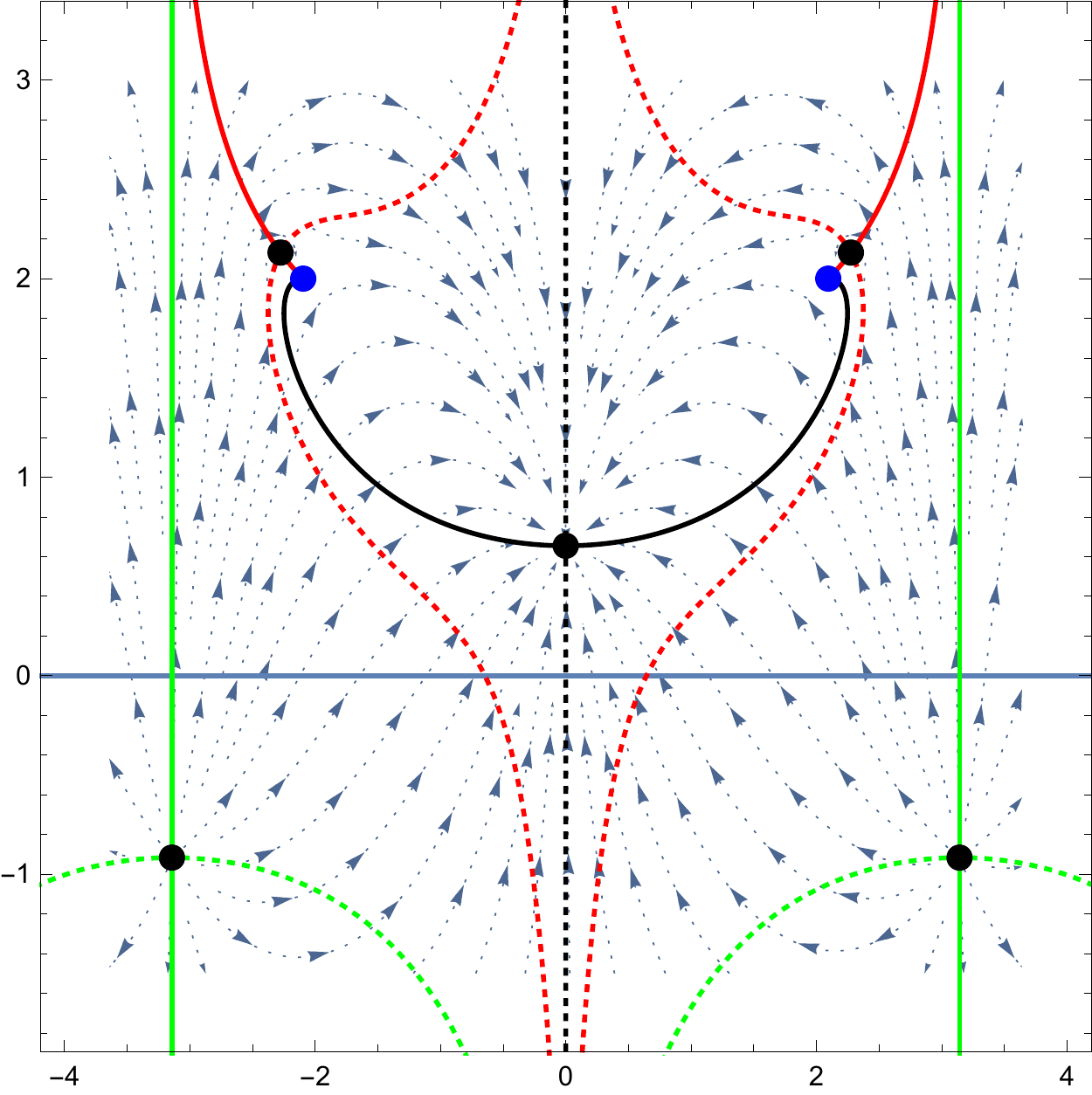}
		\subcaption{\footnotesize Contributing thimbles (full lines) and anti-thimbles 
		            (dashed lines) for the $U(1)$
		            one-link model. The critical points of the action are represented by the black dots 
		            and poles of the holomorphic flow are indicated by blue dots.
		            The contributing thimbles all end in poles at finite flow times $\tau$. 
		            The blue arrows in the background represent the drift
		            force in the Complex Langevin evolution. Due to the periodicity of 
		            the model the non-contributing thimbles (green) at the edges of the 
		            original integration domain $[-\pi, \pi]$ coincide.}
	\end{subfigure}
	\hspace*{20pt}
	\begin{subfigure}[c]{.48\textwidth}
		\includegraphics[width=.9\textwidth]{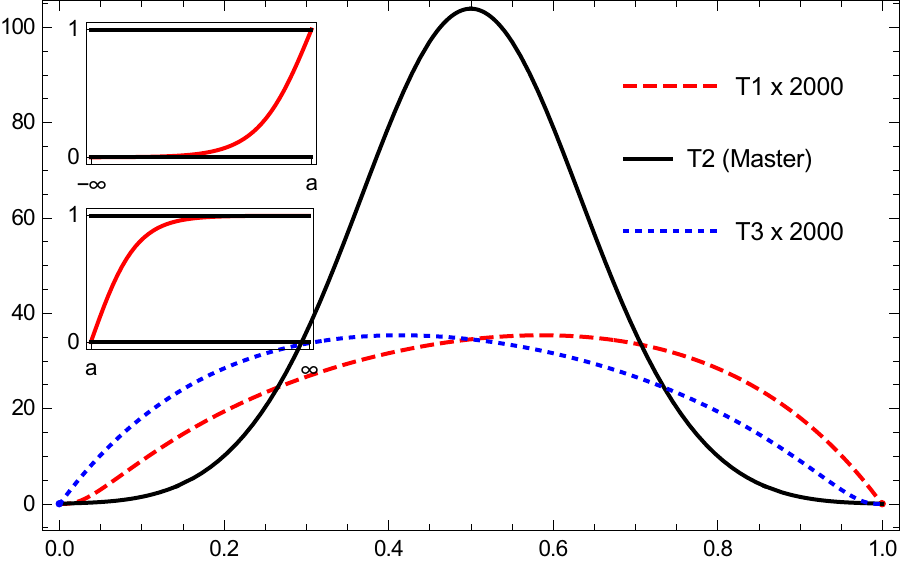}
	  \subcaption{\footnotesize Boltzmann factors 
	              $e^{-\mathrm{Re}[S_{\sigma}(\tau)]}$ vs.~the flow parameter
	  $\tau$ for the three contributing thimbles in the $U(1)$ one-link model. 
	  The black line represents the thimble with the largest weight
	  -- the master thimble used to determine
	  the relative weights. To maximize overlap of the Boltzmann distributions, the different
	  integration ranges of the three thimbles are mapped to the 
	  same interval $[0, 1]$. With $a, b \in \mathbb{R}^{+}, a \neq b$ 
	  for the two symmetric thimbles on the sides T1 and T3 (red) the parameter ranges are
	  of the form $(-\infty, a]$ and $[-a, \infty)$.
	  Suitable mappings are hyperbolic tangent 
	  functions, see the inlays. 
	  A linear transformation is used for the central thimble T2 (black) with parameter range of the
	  form $[-b, b]$. Futher remarks on the transformations can be found in \cite{Bluecher:2018sgj}
	  and the appendix B therein. 
	  For better visuability the distributions have been rescaled.}
	\end{subfigure}
	\caption{Thimble structure (a) and distribution of the Boltzmann weights for every 
	         contributing thimble (b) mapped to the interval $[0, 1]$.}
	\label{fig:U1onelink}
\end{figure}

\noindent In this section we put to work our reweighting algorithm by using 
a $U(1)$ one-link model with a finite chemical potential $\mu$
\begin{eqnarray}
Z & = & \int_{-\pi}^{\pi} dx \ e^{-S(x)}\,, \nonumber \\
S(x) & = & - \beta \cos(x) - \log(1 + \kappa \cos(x - i \mu))\,.
\label{eq:U1onelinkmodel}
\end{eqnarray} 
Despite its simplicity
the model shares general features with more complicated 
theories such as QCD at finite chemical potential.  
In the following we consider the case $\beta = 1, \kappa = 2$ and $\mu = 2$. 
The model has been studied using the Complex Langevin evolution 
in \cite{Aarts:2008rr}. Its thimble structure has been investigated in 
\cite{Tanizaki:2017yow, Aarts:2014nxa}.
%For this parameter the complex Langevin evolution and the Lefschetz thimbles have been
%compared in \cite{Aarts:2014nxa}.
There are three contributing thimbles, see the red and black full lines
in Fig.~\ref{fig:U1onelink} (a).
The two red ones are related by symmetry and their partition functions are complex 
conjugates \cite{Aarts:2014nxa}.
The second term in the effective action in (\ref{eq:U1onelinkmodel}) modeling 
a fermion determinant generates poles in the holomorphic flow (and equivalently 
in the Complex Langevin drift) in which the contributing thimbles end for finite
flow time $\tau$. 
Due to the presence of these poles the Complex Langevin evolution yields 
incorrect results for the given parameters above \cite{Aarts:2014nxa}.
For the partition functions (\ref{eq:single-th-rpf}) associated 
with the contributing thimbles 
all integration ranges for the parameter $\tau$ differ. 
This demands to transform these to the same interval
yielding overlap of the distributions $e^{-\mathrm{Re}[S_{\sigma}(\tau)]}$
which facilitates the reweighting procedure for the 
relative weights (\ref{eq:ratio}).
The remapped distributions are shown in Fig.~\ref{fig:U1onelink} (b) where details on the 
transformation mappings are provided in the caption.
The central thimble (black) carries
the largest weight. We choose it to be the master thimble. 
For a quantitative test of our method we compute the analogue of the 
Polyakov loop, its inverse, the plaquette and the density which are given in analytical
form in \cite{Aarts:2008rr}
\begin{eqnarray}
\langle U \rangle & = & \langle e^{ix} \rangle\,, \nonumber \\[2ex]
\langle U^{-1} \rangle & = & \langle e^{-ix} \rangle\,, \nonumber \\[2ex]
\langle P \rangle & = & \langle \cos(x) \rangle\,, \nonumber \\[2ex]
\langle n \rangle & = & \left \langle \frac{i \kappa \sin(x - i \mu)}
{1 + \kappa \cos(x - i \mu)} \right\rangle\,. 
\label{eq:U1obs}
\end{eqnarray}
The results of our Monte Carlo simulations are displayed in Table 
\ref{tab:results} in comparison with the exact results. This includes the relative weights 
for the contributing thimbles with respect to the master thimble of choice. 
%For the simulation we have measured each observable $O(10^9)$ times. The error estimates were obtained 
%by a standard Jackknife analysis. 
The results for the observables 
agree with the exact ones from \cite{Aarts:2008rr} within the statistical error as shown in
Table \ref{tab:results}. Systematic errors arise from the numerical 
parametrization of the thimble obtained by solving the holomorphic flow. For the simple integrals
considered here this error is estimated by comparing with the exact solution 
to be of the order of $10^{-6}$ which compares to the statistical error.
The ratio of partition functions appears to be influenced by these systematics, 
see Table \ref{tab:results}. 
Combining the statistical and the systematic error 
we find that all quantities agree with the exact results.
We remark that we obtained the same results
within the combined errorbars for either choosing 
the central or the left (right) thimble as the master thimble.
\begin{table}[t]
\begin{center}
\def\arraystretch{1.5}
\begin{tabular}{ |c | c | c |}
  \hline			
  $\langle \mathcal{O} \rangle $ & numerical & exact \\

\hline\hline

Re$\left<U\right>$ & $0.315217(3)$ &$0.315219$\\ \hline
Re$\left<U^{-1}\right>$ & $1.800941(3) $&$1.800939$\\ \hline
Re$\left<P\right>$ &$1.058079(3) $&$1.058079$\\ \hline
Re$\left<n\right>$ &$0.742861(1)$ &$0.742860$\\ \hline
$\left. Z_2/Z_1\right|_{T_1} \times 10^{-3}$ &$2.99378(3)$ &$2.99382$\\ \hline
$\left.Z_1/Z_2\right|_{T_2} \times 10^{4}$ &$3.34032(4)$ &$3.34022$\\ \hline
$\left. Z_2/Z_3\right|_{T_3} \times 10^{-3}$ &$2.99377(3)$ &$2.99382$\\ \hline
$\left.Z_3/Z_2\right|_{T_2} \times 10^{4}$ &$3.34026(9)$ &$3.34022$\\ \hline
\end{tabular}
\caption{Numerical results and exact values of observables for the U(1) one-link model together with the statistical errors. We have found the imaginary parts for the observables to be all consistent with zero within the 
statistical error. Possible deviations are caused by systematic errors from the numerical 
parametrization, see the main text for details.}
\label{tab:results}
\end{center}
\end{table}

%\subsection{One-site $z^4$ model}
%The action reads
%\begin{equation}
%S(x) = \frac{\sigma}{2} x^2 + \frac{\lambda}{4} x^4 + h x\,.
%\end{equation}
\section{Conclusions and outlook}
\noindent In this work we have developed a novel reweighting procedure to determine the relative 
weights in the thimble decomposition. The approach is successfully put to work 
by considering a $U(1)$ one-link model representing a toy model for a gauge theory with 
fermions at finite density. Although its thimble structure is trivial to parametrize numerically it provides a valuable testbed. Numerical results for the expectation values of observables 
obtained with the novel method agree well with 
exact results within the combined statistical and systematic errors. 
The extension to higher dimesional theories is challenging since Monte Carlo sampling 
on the thimble is required with or without having an explicit parametrization.
Nevertheless, the reweighting procedure 
straightforwardly generalizes to field theories and may be combined with existing simulation methods
for thimbles.  
Applications in progress are higher dimensional gauge theories \cite{HD-Bielefeld}.

\textit{Acknowledgements:} We thank K.~Fukushima, C.~Schmidt, A.~Rothkopf, F.~Ziesch\'{e}, the Heidelberg
Lattice group and the CLE collaboration for discussions and work on
related subjects. This work is supported by EMMI, the BMBF grant
05P12VHCTG, and is part of and supported by the DFG Collaborative
Research Centre "SFB 1225 (ISOQUANT)".  I.-O.~Stamatescu and
M.~Scherzer acknowledge financial support from DFG under
STA 283/16-2. F.P.G.~Ziegler is supported by the FAIR OCD
project.
\bibliographystyle{JHEP.bst}
\bibliography{literature}

\providecommand{\href}[2]{#2}\begingroup\raggedright\begin{thebibliography}{10}

\bibitem{deForcrand:2010ys}
P.~de~Forcrand, \emph{{Simulating QCD at finite density}}, {\emph{PoS}
  {\bfseries LAT2009} (2009) 010}
  [\href{https://arxiv.org/abs/1005.0539}{{\ttfamily 1005.0539}}].

\bibitem{Aarts:2008rr}
G.~Aarts and I.-O. Stamatescu, \emph{{Stochastic quantization at finite
  chemical potential}},
  \href{https://doi.org/10.1088/1126-6708/2008/09/018}{\emph{JHEP} {\bfseries
  09} (2008) 018} [\href{https://arxiv.org/abs/0807.1597}{{\ttfamily
  0807.1597}}].

\bibitem{Aarts:2017hqp}
G.~Aarts, K.~Boguslavski, M.~Scherzer, E.~Seiler, D.~Sexty and I.-O.
  Stamatescu, \emph{{Getting even with CLE}},
  \href{https://doi.org/10.1051/epjconf/201817514007}{\emph{EPJ Web Conf.}
  {\bfseries 175} (2018) 14007}
  [\href{https://arxiv.org/abs/1710.05699}{{\ttfamily 1710.05699}}].

\bibitem{Cristoforetti:2012su}
{\scshape AuroraScience} collaboration, M.~Cristoforetti, F.~Di~Renzo and
  L.~Scorzato, \emph{{New approach to the sign problem in quantum field
  theories: High density QCD on a Lefschetz thimble}},
  \href{https://doi.org/10.1103/PhysRevD.86.074506}{\emph{Phys. Rev.}
  {\bfseries D86} (2012) 074506}
  [\href{https://arxiv.org/abs/1205.3996}{{\ttfamily 1205.3996}}].

\bibitem{Alexandru:2015sua}
A.~Alexandru, G.~Basar, P.~F. Bedaque, G.~W. Ridgway and N.~C. Warrington,
  \emph{{Sign problem and Monte Carlo calculations beyond Lefschetz thimbles}},
  \href{https://doi.org/10.1007/JHEP05(2016)053}{\emph{JHEP} {\bfseries 05}
  (2016) 053} [\href{https://arxiv.org/abs/1512.08764}{{\ttfamily
  1512.08764}}].

\bibitem{Mori:2017pne}
Y.~Mori, K.~Kashiwa and A.~Ohnishi, \emph{{Toward solving the sign problem with
  path optimization method}},
  \href{https://doi.org/10.1103/PhysRevD.96.111501}{\emph{Phys. Rev.}
  {\bfseries D96} (2017) 111501}
  [\href{https://arxiv.org/abs/1705.05605}{{\ttfamily 1705.05605}}].

\bibitem{Alexandru:2018fqp}
A.~Alexandru, P.~F. Bedaque, H.~Lamm and S.~Lawrence, \emph{{Finite-Density
  Monte Carlo Calculations on Sign-Optimized Manifolds}},
  \href{https://doi.org/10.1103/PhysRevD.97.094510}{\emph{Phys. Rev.}
  {\bfseries D97} (2018) 094510}
  [\href{https://arxiv.org/abs/1804.00697}{{\ttfamily 1804.00697}}].

\bibitem{Bursa:2018ykf}
F.~Bursa and M.~Kroyter, \emph{{A simple approach towards the sign problem
  using path optimisation}},
  \href{https://arxiv.org/abs/1805.04941}{{\ttfamily 1805.04941}}.

\bibitem{Schmidt:2017gvu}
C.~Schmidt and F.~Ziesch\'e, \emph{{Simulating low dimensional QCD with
  Lefschetz thimbles}}, {\emph{PoS} {\bfseries LATTICE2016} (2017) 076}
  [\href{https://arxiv.org/abs/1701.08959}{{\ttfamily 1701.08959}}].

\bibitem{DiRenzo:2017igr}
F.~Di~Renzo and G.~Eruzzi, \emph{{One-dimensional QCD in thimble
  regularization}},
  \href{https://doi.org/10.1103/PhysRevD.97.014503}{\emph{Phys. Rev.}
  {\bfseries D97} (2018) 014503}
  [\href{https://arxiv.org/abs/1709.10468}{{\ttfamily 1709.10468}}].

\bibitem{Witten:2010cx}
E.~Witten, \emph{{Analytic Continuation Of Chern-Simons Theory}}, {\emph{AMS/IP
  Stud. Adv. Math.} {\bfseries 50} (2011) 347}
  [\href{https://arxiv.org/abs/1001.2933}{{\ttfamily 1001.2933}}].

\bibitem{Alexandru:2015xva}
A.~Alexandru, G.~Basar and P.~Bedaque, \emph{{Monte Carlo algorithm for
  simulating fermions on Lefschetz thimbles}},
  \href{https://doi.org/10.1103/PhysRevD.93.014504}{\emph{Phys. Rev.}
  {\bfseries D93} (2016) 014504}
  [\href{https://arxiv.org/abs/1510.03258}{{\ttfamily 1510.03258}}].

\bibitem{DiRenzo:2017omx}
F.~Di~Renzo, \emph{{Simulating lattice field theories on multiple thimbles}},
  in \emph{{35th International Symposium on Lattice Field Theory (Lattice 2017)
  Granada, Spain, June 18-24, 2017}}, 2017,
  \href{https://arxiv.org/abs/1710.06958}{{\ttfamily 1710.06958}},
  \href{http://inspirehep.net/record/1631649/files/arXiv:1710.06958.pdf}{http://inspirehep.net/record/1631649/files/arXiv:1710.06958.pdf}.

\bibitem{Bluecher:2018sgj}
S.~Bluecher, J.~M. Pawlowski, M.~Scherzer, M.~Schlosser, I.-O. Stamatescu,
  S.~Syrkowski et~al., \emph{{Reweighting Lefschetz Thimbles}},
  \href{https://doi.org/10.21468/SciPostPhys.5.5.044}{\emph{SciPost Phys.}
  {\bfseries 5} (2018) 044} [\href{https://arxiv.org/abs/1803.08418}{{\ttfamily
  1803.08418}}].

\bibitem{Tanizaki:2017yow}
Y.~Tanizaki, H.~Nishimura and J.~J.~M. Verbaarschot, \emph{{Gradient flows
  without blow-up for Lefschetz thimbles}},
  \href{https://doi.org/10.1007/JHEP10(2017)100}{\emph{JHEP} {\bfseries 10}
  (2017) 100} [\href{https://arxiv.org/abs/1706.03822}{{\ttfamily
  1706.03822}}].

\bibitem{Aarts:2014nxa}
G.~Aarts, L.~Bongiovanni, E.~Seiler and D.~Sexty, \emph{{Some remarks on
  Lefschetz thimbles and complex Langevin dynamics}},
  \href{https://doi.org/10.1007/JHEP10(2014)159}{\emph{JHEP} {\bfseries 10}
  (2014) 159} [\href{https://arxiv.org/abs/1407.2090}{{\ttfamily 1407.2090}}].

\bibitem{HD-Bielefeld}
J.~M. Pawlowski, M.~Scherzer, C.~Schmidt, F.~Ziegler and F.~Ziesch\'e, \emph{in
  preparation}.

\end{thebibliography}\endgroup

\end{document}